\documentclass[times,preprint]{aastex631}

\usepackage[T1]{fontenc}
\usepackage{epsf,color,amsmath}

\usepackage{cancel}

\newcommand{\sfrac}[2]{\mathchoice%
  {\kern0em\raise.5ex\hbox{\the\scriptfont0 #1}\kern-.15em/
    \kern-.15em\lower.25ex\hbox{\the\scriptfont0 #2}}
  {\kern0em\raise.5ex\hbox{\the\scriptfont0 #1}\kern-.15em/
    \kern-.15em\lower.25ex\hbox{\the\scriptfont0 #2}}
  {\kern0em\raise.5ex\hbox{\the\scriptscriptfont0 #1}\kern-.2em/
    \kern-.15em\lower.25ex\hbox{\the\scriptscriptfont0 #2}} {#1\!/#2}}

\newcommand{\castro}{{\sf Castro}}

\newcommand{\amrex}{{\sf AMReX}}

\newcommand{\isotm}[2]{{}^{#2}\mathrm{#1}}

\newcommand{\Ub}{\mathbf{U}}

\newcommand{\omegadot}{\dot{\omega}}
\newcommand{\Sdot}{\dot{S}}

\usepackage{bm}

\newcommand{\Uc}{{\,\bm{\mathcal{U}}}}

\begin{document}
\title{Practical Effects of Integrating Temperature with Strang Split Reactions}

\shorttitle{Temperature Integration with Strang}

\shortauthors{}

\author[0000-0001-8401-030X]{M.~Zingale}
\affiliation{Dept.\ of Physics and Astronomy, Stony Brook University,
       Stony Brook, NY 11794-3800}

\author[0000-0003-0439-4556]{M.~P.~Katz}
\affiliation{NVIDIA Corporation}

\author[0000-0003-2300-5165]{D.\ E.\ Willcox}
\affiliation{Lawrence Berkeley National Laboratory, Berkeley, CA}

\author[0000-0002-1530-781X]{A.\ Harpole}
\affiliation{Dept.\ of Physics and Astronomy, Stony Brook University,
             Stony Brook, NY 11794-3800}

\begin{abstract}
For astrophysical reacting flows, operator splitting is commonly used
to couple hydrodynamics and reactions.  Each process operates
independent of one another, but by staggering the updates in a
symmetric fashion (via Strang splitting) second order accuracy in time
can be achieved.  However, approximations are often made to the
reacting system, including the choice of whether or not to integrate
temperature with the species.  Here we demonstrate through a simple
convergence test that integrating an energy equation together with reactions
achieves the best convergence when modeling reactive flows with Strang
splitting.  Additionally, second order convergence cannot be achieved without integrating an energy or temperature equation.
\end{abstract}

\keywords{hydrodynamics---methods: numerical}

\section{Introduction}\label{Sec:Introduction}

Simulations of stellar flows require solving the equations of
hydrodynamics coupled with a nuclear reaction network.  The equations
of hydrodynamics with reacting sources are:
\begin{align}
\frac{\partial \rho}{\partial t} + \nabla \cdot (\rho \Ub) &= 0 \\
\frac{\partial (\rho X_k)}{\partial t} + \nabla \cdot (\rho X_k \Ub) &= \rho \omegadot_k \\
\frac{\partial (\rho \Ub)}{\partial t} + \nabla \cdot (\rho \Ub \Ub) + \nabla p &= 0 \\
\frac{\partial (\rho E)}{\partial t} + \nabla \cdot \left [ (\rho E + p) \Ub \right ] &= \rho \Sdot
\end{align}
where $\rho$ is the density, $\Ub$ is the velocity vector, $X_k$ are the species mass fractions
with creation rates $\omegadot_k$, $p$ is the pressure, $E$ is the
specific total energy, and $\Sdot$ is the nuclear energy generation rate.

When we are reacting, we can look at internal energy
\begin{equation}
\rho \frac{De}{Dt} + p \nabla \cdot \Ub = \rho \Sdot,
\end{equation}
where $e$ is the specific internal energy or alternately, we can evolve the temperature, $T$,
\begin{equation}
\rho c_v \frac{DT}{Dt} = \rho \left (\frac{p}{\rho^2} - \left .\frac{\partial e}{\partial \rho} \right |_T \right ) \frac{D\rho}{Dt} + \rho \Sdot,
\end{equation}
where $c_v = \partial e / \partial T |_\rho$ (this form neglects composition
changes, see \citealt{ABNZ:III}).

\section{Numerical Methodology}

We use the freely-available \castro\ simulation
code~\citep{castro,castro_joss} to solve the equations of
hydrodynamics, using an unsplit piecewise parabolic method coupled with
reactions.  \castro\ uses either Strang splitting or
spectral deferred corrections (SDC) to couple the hydrodynamics and
reactions~\citep{castro_sdc}.  Here we focus on the Strang splitting.

In a Strang split evolution~\citep{strang:1968}, we update the full
hydrodynamics state, $\Uc$, as:
\begin{equation}
  \Uc^{n+1} = {\bf R}_{\Delta t/2} {\bf A}_{\Delta t} {\bf R}_{\Delta t/2} \Uc^n
\end{equation}
where ${\bf R}_{\Delta t/2}$ is the reaction update through a timestep
$\Delta t/2$ and ${\bf A}_{\Delta t}$ is the advective update through
$\Delta t$.  We see with this splitting, each process operates on the
state left behind by the previous operation, and the staggering of the
physics is done to give second order accuracy in time.

During reactions, we neglect the hydrodynamics terms,
so the reactive system updates according to:
\begin{align}
\frac{D\rho}{Dt} &= 0 \\
\frac{DX_k}{Dt} &= \omegadot_k \label{eq:species}
\end{align}
with either
\begin{equation}
\frac{DT}{Dt} = \frac{\Sdot}{c_v}
\end{equation}
or
\begin{equation}
\frac{De}{Dt} = \Sdot
\end{equation}

There are several different approaches taken in the literature to
this operator-split reacting system, including some approximations
that make integrating the reaction system computationally less expensive:
\begin{itemize}
\item Evolve $(X_k)$ only.  This neglects temperature evolution
  completely in the burn, only evolving Eq.~\ref{eq:species}.  This is
  the method used in \citet{flash}.

\item Evolve $(X_k, T)$.   This is used in \cite{Pakmor:2012} and
  \cite{Garcia-Senz:2013}.  To avoid expensive equation of state calls
  in getting the specific heat, we can optionally ``freeze'' the value
  of $c_v$ at the start of integration.  This was discussed in \cite{Bell:2004} and
  until recently was the default method in \castro.

\item Evolve $(X_k, e)$, and get $T$ from $e$ using the equation of
  state.  This was discussed in \citet{fma} and is the current default
  method in \castro.
\end{itemize}
\cite{Raskin:2010} also propose a hybrid system where the first approach is
used in most cases, switching to the second approach only near NSE.
For all of these cases, density is constant during the reaction operation.
Depending on how vigorous the burning is and how much the temperature
changes during a hydrodynamic timestep, one or more of these
methods may be reasonable.  For explosive reactions, we expect
that evolving the full system will be required.  The goal of this
note is to try to quantify the convergence of a reacting hydrodynamics
code with these different approximations.

In \citet{castro_sdc}, we introduced a test problem where we could
measure the convergence of a reacting hydrodynamics problem via
Richardson extrapolation---this was an acoustic pulse with helium
burning via $3$-$\alpha$ and
$\isotm{C}{12}(\alpha,\gamma)\isotm{O}{16}$.  Initially, the domain is
pure $\isotm{He}{4}$, but both $\isotm{C}{12}$ and $\isotm{O}{16}$ are
created as time evolves.  The published tests 
showed that we can get overall 4th order in space and time
convergence with SDC coupling.  Here we run the same test with Strang coupling, looking
at the various approaches at incorporating a temperature / energy equation in
the reactive portion of the update.

For each method, we run the \castro\ {\tt reacting\_convergence} test
problem at 5 resolutions: $64^2$, $128^2$, $256^2$, $512^2$, and
$1024^2$, with the timestep kept fixed in proportion to the grid resolution.
We then compute four errors between adjacent resolutions by
coarsening the finer resolution run, and computing the $L_1$ norm over
all zones.  

\begin{figure}[t]
\centering
\plotone{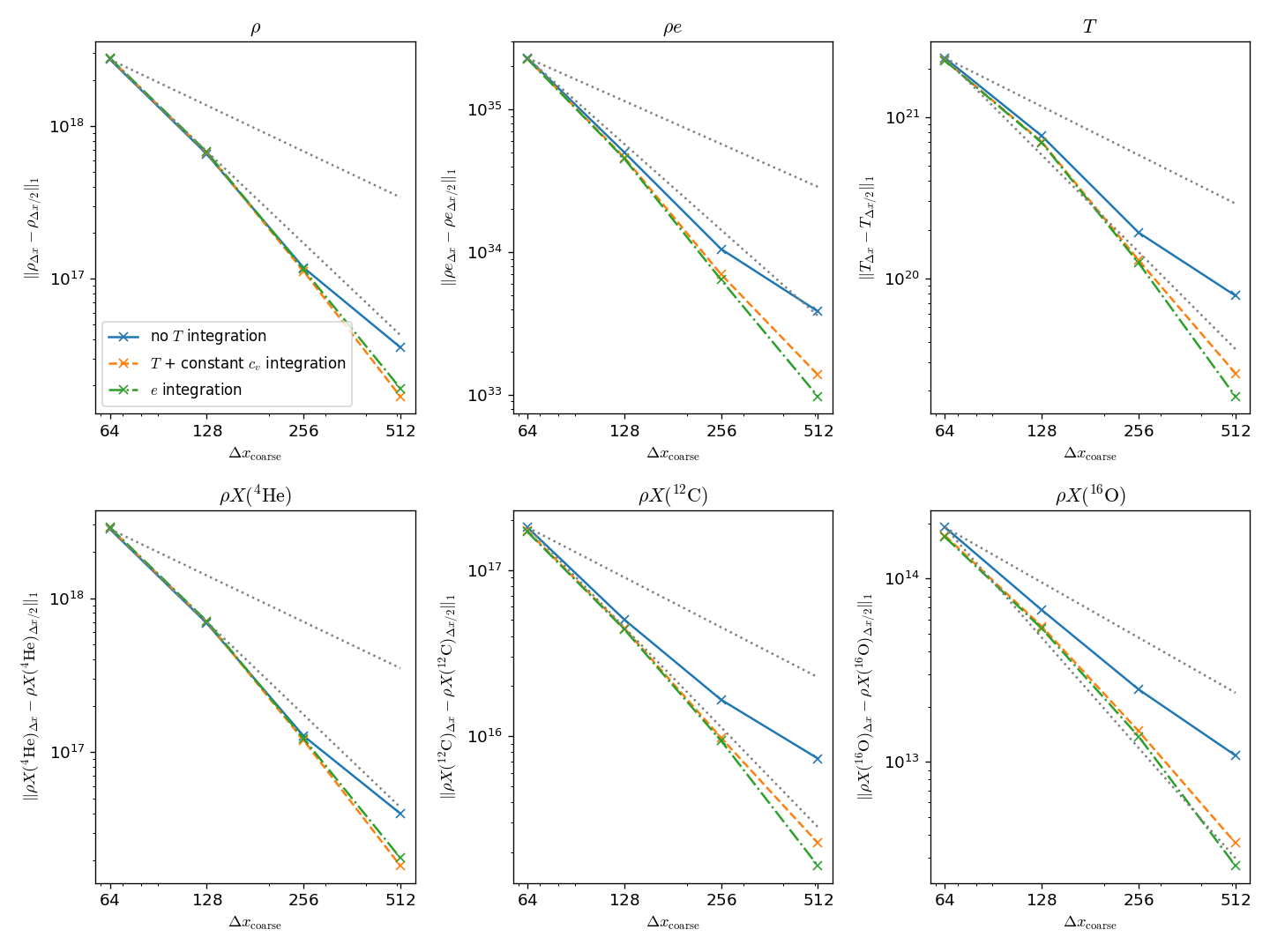}
\caption{\label{thefigure} Convergence of fluid quantities as a function of resolution
for 3 different Strang equation systems: just evolving $X_k$, evolving $(X_k, T)$ with $c_v$ held fixed, and evolving $(X_k, e)$. The dotted lines show ideal first and second order convergence.}
\end{figure}

Figure~\ref{thefigure} shows the norm of the error vs.\ the coarse run
resolution.  The slope of these lines is a measure of the convergence
rate~\citep{ORVV}.  We see that all methods converge at least second order
for density, but for the thermodynamic quantities, $(\rho e)$ and $T$,
the method where only $X_k$ is evolved during reactions has larger
errors and much worse convergence than the other methods.  Looking at
the trace nuclei generated in the burning, $\isotm{C}{12}$ and
$\isotm{O}{16}$, we see a large difference between the two methods
that evolve some sort of energy and the one method where only $X_k$ is
evolved---the latter converging essentially first order at high
resolution.

\section{Summary}

Looking at global convergence of a reacting hydrodynamics problem we
see that second order convergence is only realized when temperature or
energy is evolved alongside the composition when using a Strang-split
approach to reactions.  This is just a single, rather simple problem,
but this suggests that reactive hydrodynamics simulations should
switch to integrating temperature or another energy equation together
with reactions to yield better overall convergence and accuracy.  This
complements the work of \cite{muller:1986} which showed that when
evolving near nuclear statistical equilibrium, evolving entropy
with the system is needed for stability.  We expect that for problems
with vigorous burning, such as detonations, directly coupling the composition
and thermodynamic evolution will be especially important.

\begin{acknowledgments}

\castro\ is freely available at
\url{https://github.com/AMReX-Astro/Castro}.   The work at
Stony Brook was supported by DOE/Office of Nuclear Physics grant
DE-FG02-87ER40317.  This material is based upon work supported by the
U.S. Department of Energy, Office of Science, Office of Advanced
Scientific Computing Research and Office of Nuclear Physics,
Scientific Discovery through Advanced Computing (SciDAC) program under
Award Number DE-SC0017955.  This research was supported by the
Exascale Computing Project (17-SC-20-SC), a collaborative effort of
the U.S. Department of Energy Office of Science and the National
Nuclear Security Administration.

\end{acknowledgments}

\software{\amrex\ \citep{amrex_joss},
          \castro\ \citep{castro_joss},
          matplotlib \citep{Hunter:2007},
          NumPy \citep{numpy},
          VODE \citep{vode}
         }


\bibliographystyle{aasjournal}
\bibliography{ws}

\end{document}